\documentclass[a4paper]{article}

\usepackage[english]{babel}
\usepackage[utf8x]{inputenc}
\usepackage{amsmath}
\usepackage{graphicx}
\usepackage{pifont}
\usepackage{epsfig}
\usepackage{todonotes}
\newcommand{\cmark}{\color{green} \ding{51}}
\newcommand{\xmark}{\color{red} \ding{55}}
\providecommand{\keywords}[1]
{
  \small	
  \textbf{\textit{Keywords---}} #1
}

\title{A Privacy-Preserving Solution for Proximity Tracing Avoiding Identifier Exchanging}

\author{Francesco Buccafurri$^{1}$, Vincenzo De Angelis$^{1}$, Cecilia Labrini$^{1}$  \\
        \small $^{1}$DIIES, University Mediterranea of Reggio Calabria, Reggio Calabria, Italy \\
        \small \{bucca,vincenzo.deangelis,cecilia.labrini\}@unirc.it \\
}

\begin{document}

\maketitle
\begin{abstract}
Digital contact tracing is one of the actions useful, in combination with other measures, to manage an epidemic diffusion of an infection disease in an after-lock-down phase. This is a very timely issue, due to the pandemic of COVID-19 we are unfortunately living. Apps for contact tracing aim to detect proximity of users and to evaluate the related risk in terms of possible contagious. Existing approaches leverage Bluetooth or GPS, or their combination, even though the prevailing approach is Bluetooth-based and relies on a decentralized model requiring the mutual exchange of ephemeral identifiers among users' smartphones. Unfortunately, a number of security and privacy concerns exist in this kind of solutions, mainly due to the exchange of identifiers, while GPS-based solutions (inherently centralized) may suffer from threats 
concerning massive surveillance.
In this paper, we propose a solution leveraging GPS to detect proximity, and Bluetooth only to improve accuracy, without enabling exchange of identifiers.
Unlike related existing solutions, no complex cryptographic mechanism is adopted, while ensuring that the server does not learn anything about locations of users.

\end{abstract}


\section{\uppercase{Introduction}}
\label{sec:introduction}

The epidemic diffusion of an infection disease can be contrasted by adopting various actions, suitably combined each other, like tests,  pharmacological treatments, 
quarantine, contact tracing.
The latter consists in identifying the maximum number of persons potentially infected by a given patient detected as positive to the infection, in a suitable contagious temporal (past) window.
Which kind of contact makes a person potentially infected and how long the contagious window is, strictly depend on the type of infection.
The pandemic of COVID-2019 we are leaving in this period is characterized by high contagiousness mainly due to asymptomatic and pre-symptomatic infections, during a large temporal window (at least 14 days) \cite{cheng2020contact}.
Therefore, traditional contact tracing based on human intelligence activities to identify contacts is not sufficient if not supported by digital solutions able to capture even short (numerous) contacts occurred during the activities of daily life \cite{ferretti2020quantifying}.
For this reason, there is nowadays a great attention towards digital contact tracing that many countries in the world are adopting through mobile apps to better manage the after-lock-down phase.

Bluetooth Low Energy (BLE) \cite{gomez2012overview} on board of smartphones is the technology used to implement decentralized protocols, in which users in BLE action range exchange pseudonym identities and store them together with some information useful to evaluate the risk of the occurred contacts in terms of possible contagious.
The DP-3T based solutions \cite{troncosodecentralized, apgo} fall in the above category and certainly represent the prevailing approach in this moment, recognized as the approach that best protects citizens' privacy.

However, DP-3T is not immune from threats to privacy and to the integrity of the protocol, also due to some technological issues related to its Bluetooth-based implementation \cite{avitabile2020towards,dp3tblevulnerability,vaudenay2020centralized}.

On the other hand, centralized solutions are often based on GPS.
The basic way to implement a GPS-based solution requires that the user's absolute position is periodically transmitted to a server
(under the control of the government), which is then able to maintain the graph of contacts, possibly in form of pseudonyms. 
One of the advantages of the centralized model is that identities are not exchanged among users, and this disarm a number of possible issues arising from the possible misbehaviour of users.
However, as recently stated by EU \cite{EU2020}, GPS-based solutions introduce intolerable risks threatening fundamental rights of people, if implemented as above,
in the general case that the server cannot be assumed fully trusted
and positions of user are stored or potentially inferred.

On the other hand, more sophisticated approaches relying on GPS exist, which, thanks to multi party computation and other complex cryptographic mechanisms, are able to effectively contrast the issues arising from non-trusted servers \cite{reichert2020privacy,berke2020assessing}.
However, these solutions are not scalable \cite{dar2020applicability}, due to the computational overhead required by cryptographic protocols.

In this paper, we propose a solution,
called \textit{Zero Ephemeral Exchanging-Privacy-Preserving-Proximity Tracing} (ZE2-P3T, for short),
relying on GPS to detect proximity,
and on Bluetooth (BLE) only to improve accuracy.
Our solution overcomes the most security and privacy issues of the
DP-3T approach, basically because the exchange of identifiers
is not enabled.
Interestingly, unlike related existing solutions, 
to ensure that the server does not learn anything about locations of users,
no complex cryptographic mechanism is adopted, making our solution feasible also for a large number of users.

The structure of this paper is the following. 
In Section \ref{sec:related_work}, the related literature is analysed.
Section \ref{sec:background} describe the state-of-the-art decentralized protocol DP-3T is described and the motivation of our study. In Section \ref{sec:proposed_model}, we present ZE2-P3T, a new solution which does not require the exchange of identifiers between the users. ZE2-PT3 uses a protocol called PNP, described in Section \ref{sec:pnp}, to improve the accuracy of the GPS localization. The security analysis is discussed in Section \ref{sec:security_analysis}.
Finally, in Section \ref{sec:conclusion}, we draw our conclusions.

\section{\uppercase{Related Work}} \label{sec:related_work}

The COVID-19 pandemic certainly represents one of the most difficult challenges that modern society has ever faced. To counter and slow down the spread of the virus, new ways, new strategies, and solutions are being sought every day, in every sector, from the economic to the medical, from the political to the technological. Precisely in this latter field, researchers from all over the world are investing their energies to propose, as quickly as possible, digital solutions for tracking contacts that preserve privacy and that comply with current regulations. 

Many solutions decide upon for a Bluetooth-based approach. Several solutions opt for a decentralized approach (such as DP-3T and the Apple and Google Exposure Notification System) \cite{avitabile2020towards} to guarantee high privacy properties. Among the decentralized models, the emerging model is certainly the Decentralized Privacy-Preserving Proximity Tracing (DP-3T) \cite{troncosodecentralized} and, therefore, the DP-3T based solutions. This model is based on ephemeral pseudonyms (called EphID) sent via Bluetooth Low Energy (BLE) which are registered by nearby users. We will see more carefully this model in the next section. In the spirit of collaboration, Google and Apple announced a joint effort for a new Bluetooth protocol that preserves privacy to support Exposure Notification \cite{apgo} and which will follow the DP-3T principles. Avitabile et al. \cite{avitabile2020towards} unveiled Pronto-C2, a decentralized tracking system that is based on BLE and appears to be more resistant than DP-3T against mass surveillance attacks. This system can be implemented through government servers but it can also be completely decentralized by using blockchain technology. CAUDHT is a decentralized system based on Distributed Hash Tables \cite{brack2020decentralized}. The entities involved are users, a distributed hash table (DHT), and the Health Authority (HA), which has the role of confirmation of positive cases. Another decentralized protocol based on Bluetooth is TCN (Temporary Contact Numbers) \cite{tcn}. To solve the problem of scalability, the protocol switches from purely random TCNs to TCNs generated deterministically from some seed data. The price it pays for greater scalability is a reduction in privacy because the TCNs derived from the same report can be linked together. 

Other solutions choose a centralized approach \cite{avitabile2020towards} such as NTK and ROBERT which have been developed inside the Pan-European Privacy-Preserving Proximity Tracing (PEPP-PT) \cite{pepp}. Centralization has the advantage of providing epidemiologists with more useful data, thus allowing more effective actions to be taken to defeat the virus, but some scholars fear that these systems could become a more intrusive massive surveillance tool for governments \cite{vaudenay2020centralized}. PEPP-PT NTK is a proximity tracing system, based on Bluetooth Low Energy \cite{ntkpepp}. Just like DP-3T, NTK and ROBERT are based on ephemeral pseudonyms sent via BLE that are registered by nearby users, with the difference that the secret keys for calculating EphIDs are created and managed by a back-end server and not from the user's phone \cite {aisec2020pandemic}. Moreover, DP-3T, NTK, and ROBERT require a central backend for their operation in which epidemiological data is stored. 
The Altuwaiyan et al. model, called EPIC, \cite{altuwaiyan2018epic} is always based on Bluetooth technology, and offers a fine-grained human-to-human contact tracing scheme with hybrid wireless and localization technology. EPIC introduces a matching method which uses homomorphic encryption to match common wireless devices between the infected and the regular user. However, the system can suffer from serious privacy losses and above all, it has scalability problems \cite{dar2020applicability}.

\textit{TraceTogether}, implemented in Singapore,  was the first centralized 
Bluetooth-based solution \cite{singaporean}. This approach involves two types of entities namely the Ministry of Health (MoH), which all users are assumed to trust, and users (who are not required to share everything with MoH if they have not been in close contact with a confirmed positive). This system manages to trace the COVID-19 transmission graph in the population that installed the app. However, solutions based on Bluetooth technology present several vulnerabilities \cite{blevul}  and have a slow and low adoption rate which therefore limits the user base adhering to the system \cite{dar2020applicability}. 
Furthermore, these systems only consider the human-to-human interaction and therefore do not allow us to discover the place where an outbreak has developed as it is not possible to know the user's position. The side effect is that such solutions ignore the fact that the COVID-19 can also be transmitted through common environments or commonly touched surfaces \cite{kampf2020persistence} (\textit{indirect} contacts). Obviously, to know the location of a user would have an intolerable price in terms of privacy.

However, trickier approaches relying on GPS localization exist, which are able to mitigate the privacy issues mentioned above.
Berke et al. \cite{berke2020assessing} propose a GPS-based solution that takes advantage of the partitioning of fine-grained GPS positions and private set intersection that allows the system to detect when a user approached positive patients. Reichert et al. \cite{reichert2020privacy} offer a solution on how to make contact tracking centralized based on GPS data to preserve user privacy. The system uses a central party (HA) and applies multi-party computation (MPC) on the real-world problem of centralized contact trace. 

However, these solutions are not scalable \cite{dar2020applicability}, due to the computational overhead required by the adopted cryptographic protocols
(i.e., MPC).

Our solution starts from the above reference framework, with the aim to overcome the privacy issues of decentralized solutions, on the one hand, and the scalability problems of centralized (absolute-position based) approaches, on the other hand.
Our approach is centralized and is based on privacy-preserving absolute position detection. The position is obtained by using GPS, in combination with BLE and the Earth magnetic field for the indoor environments.
This choice is supported by results available in the literature like \cite{de2014indoor}, which presents a system  able to 
guarantee a maximum positioning error of less than 10 cm in an internal environment.

\section{\uppercase{Background and Motivations}}\label{sec:background}

As mentioned in the previous section, the DP-3T protocol \cite{troncosodecentralized} represents at moment the prevailing approach
especially in European Union.
Despite the fact that DP-3T, similarly to TCN \cite{tcn},
suffers from some serious drawbacks concerning users' privacy,
it is the reference approach because is the state-of-the-art implementation of the decentralized model, which is preferred to the centralized model.

It is then important to describe into detail how DP-3T solutions work.
The basic idea is to install an app on each smartphone and to use Bluetooth to interact with other nearby smartphones to register the contacts. Therefore, the actors of the DP-3T system are:
\begin{itemize}
\item The \textit{users} in possession of a communication device (a smartphone equipped with Bluetooth running the DP-3T app).
\item The \textit{back-end server}, which acts exclusively as a communication platform and does not perform any processing. Moreover, the privacy of users in the system does not depend on the actions of this server.
\item The \textit{health authority}, which is responsible for informing patients of the positive test results, allows uploads from phones to the back-end and determines the contagious window.
\end{itemize}
The app broadcasts an ephemeral pseudo-random ID that represents the user and also records pseudo-random IDs observed by smartphones in the immediate proximity. If a user finds out that she/he is positive for COVID-19 then, after obtaining the approval of the health authority, may upload some anonymous data from her/his smartphone to a central server. Before uploading, all data remains exclusively on the user’s smartphone. The DP-3T model provides two decentralized proximity tracing designs: the first, defined as {\em low-cost}, is a lightweight system at the cost of limited tracing of infected patients, the second, defined as {\em unlinkable}, offers extra privacy properties with a small increase in bandwidth. 
The first solution reveals minimal information to the back-end server. Each smartphone generates an initial random initial daily key  $SK_t$ for the current day $t$ and, every day rotates the secret day key $SK_t$  by calculating $SK_t  = H (SK_{t-1})$, where $H$ is a cryptographic hash function. Each smartphone uses the secret key $SK_t$  during the day $t$  to locally generate a list of ephemeral identifiers $(EphID)$s  that change frequently (every epoch with length $l$). Therefore, at the beginning of each day $t$, each smartphone generates locally a list of $n = (24 \cdot 60) / l$ new $EphID_i$s to be transmitted during the day $t$. Given the secret day key $SK_t$, each device calculates $EphID_1 || ... || EphID_n = PRG (PRF (SK_t, broadcast key))$, where PRG is a stream cipher, PRF is a pseudo-random function, and \textit{broadcast key} is a fixed and public string. The $EphID_i$s are transmitted in random order and each $EphID$ is transmitted for $l$ minutes. The  $EphIDs$ are broadcasted via Bluetooth Low Energy announcements (the system relies on BLE beacons, whose payload is of 16 bytes, which technically limits the size of the $EphID$s).  
These $EphID$s are then locally stored (together with the corresponding proximity, the duration and an approximate indication of the time) by the other nearby smartphones. Each smartphone stores the $SK$ keys it has generated in the last 14 days and the same happens for all the data and the $EphID$s observed and generated. A user who tested positive, only after obtaining authorization from the health authority, may send to the back-end the key $SK_t$ and the day $t$ corresponding to the first day on which it was considered contagious. The back-end collects the pairs ($SK_t$, $t$) of the infected patients and periodically sends them to all the other smartphones in the system. Given the key $SK_t$, everyone can calculate all the ephemeral identifiers $EphID$s used by the infected patient starting from the corresponding day $t$. The device determines the owner’s risk score using the risk-scoring algorithm with its local records corresponding to the infectious $EphID$.

The second solution, defined as {\em unlinkable}, offers better privacy properties at the cost of a greater volume of downloads and storage space required by the smartphone. The ephemeral identifiers of positive individuals are hashed and stored in a Cuckoo filter \cite{fan2014cuckoo}, which is distributed to the users of the system. The smartphone draws a random 32-byte per-epoch seed ($seed_i$) and generate the ephemeral Bluetooth identifier $EphID_i$ for each epoch $i$: $EphID_i= TRUNCATE128( H( seed_i ) )$, where $H$ is a cryptographic hash function, and $TRUNCATE128$ truncates the output to 128 bits. Smartphones store the seeds corresponding to all past epochs in the last 14 days. For each observed $EphID$, the smartphone stores the hashed string $H (EphID || i)$, the proximity, the duration, and an approximate indication of the time. Unlike the previous solution, when a user tested positive, she/he can choose the set of epochs $I$ for which she/he wants to reveal her/his identifiers, that is, she/he can selectively decide which identifiers she/he wants to communicate to the server. After making this decision, the smartphone loads the set of pairs ($i, seed_i $). Periodically, the back-end creates a new Cuckoo filter $F$ and, for each pair ($i, seed_i$) loaded by an infected user, inserts $H (TRUNCATE128 (H (seed_i)) || i)$ into the Cuckoo filter F and sends this filter to all the smartphones in the system. Each smartphone uses this filter to determine whether the user has been in contact with an infected person.

DP-3T  suffers from several attacks that can compromise user privacy and potentially lead to undergo undetectable mass surveillance attacks \cite{avitabile2020towards}.  Some attacks may have the purpose of building a mass surveillance infrastructure to track citizens both in the case of trusted servers and in the case of colluding server (Paparazzi and Orwell attack, respectively). Another attack could aim to create a mapping between a user's real identity and her/his pseudonym (Brutus Attack). Another goal of an attack is to produce plausible digital evidence of an encounter with an infected user (Gossip Attack). An attacker could also produce false alarms so that non-at-risk user is mistakenly alerted and declared positive (Matteotti Attack). The attacks mentioned here will be thoroughly described in Section \ref{sec:security_analysis}, but we introduce this issue here because it is the consideration from which we start as a motivation of our paper.
In fact, our paper tries to offer a new declination of the centralized model overcoming the security and privacy drawbacks of DP-3T,
without introducing risks usually associated with centralized digital contact tracing at feasible computational cost for the server.

\begin{figure}[t]
        {\includegraphics[width=6.5cm]{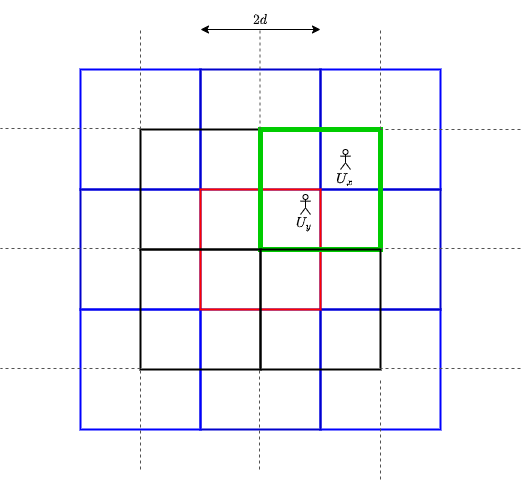}}
        \centering
        \caption{Coverage of microcells}
        \label{fig:Cells}
\end{figure}

\section{\uppercase{The Zero Ephemeral Exchanging-Privacy-Preserving-Proximity solution}} \label{sec:proposed_model}

In this section, we describe ZE2-P3T, which is the solution we propose for proximity tracing not relying on the exchange of ephemeral identifiers like
the state-of-the-art solutions.

We refer to a large geographic area $A$ representing, for example, a country.
In our model, $A$ contains several \textit{microcells} $c_i$ such that
\begin{itemize}
    \item They cover all the area $A$ 
    \item If the distance between two users $U_x$ and $U_y$ is less than a threshold parameter $d$, it exists a microcell which contains both $U_x$ and $U_y$.
\end{itemize}
 
Microcells are squares of side $2d$ organized as in Figure \ref{fig:Cells}.

Therein, we use different colours to better highlight the different microcells (13 in total). It easy to see that a user is always, simultaneously, in two different microcells and that two users positioned closest than the distance $d$ each other have a microcell in common. For example, in the figure, the user $U_x$ is in the blue and green microcells while the user $U_y$ is in the red and green microcells.

With each microcell $c_i$, we associate a point $C_i$ called \textit{centroid} corresponding to the centre of the square. The set of all the centroids is public and each user, through the combination of GPS and magnetic position systems \cite{de2014indoor}, for indoor positions,
is  able  to  identify  the  centroids  associated with the two microcells  where  the  user  is  located. 
The exact utilization of  magnetic positioning is out of the scope of this paper, even though the state-of-the-art technologies can be directly used for our purpose.
From now on, for simplicity, we refer only to the GPS signal.

Our solution requires the collaboration of a telephone service provider TSP which, periodically, sends a random $R_P$ to all users in a fixed area $Q$ (containing several microcells) according to the coverage range of the antennas. For each area $Q$, a different $R_P$ is sent by TSP and it is important that each microcell is entirely contained in an area $Q$, so that two users in the same microcell receive always the same random $R_P$ at the same time. We assume that this service is provided by a unique TSP
(to avoid complex coordination of multiple TSP in overlapping cells), and that the roaming mechanism can be enabled to ensure the maximum coverage.

\begin{figure}[t]
        \centering
        {\includegraphics[width=8cm]{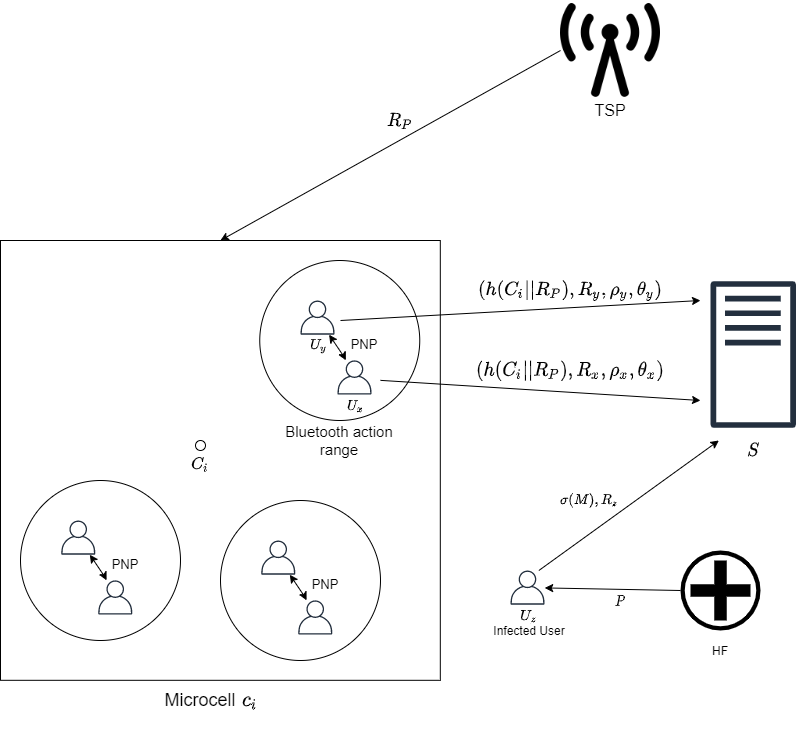}}
        \caption{The ZE2-P3T Architecture.}
        \label{fig:fileSharing}
\end{figure}

As explained below, $R_P$ is used to avoid dictionary-based attacks in order to locate the position of the user.

To guarantee the privacy, each user $U_x$, with a certain frequency, generates a random $R_x$ which is pseudonym identifier valid until a new random is generated. However, when $R_x$ expires, it is not burnt by $U_x$, but it is stored for some time.

For each of the two microcells $c_i$ where $U_x$ is located, she/he recovers the position of the centroid $C_i$ of $c_i$ and the pair $({\rho}_x, {\theta}_x)$ which represents the polar coordinates of the position of $U_x$ respect to $C_i$.
Since the GPS accuracy is not sufficient for our purpose, when $U_x$ comes in contact with another user, they exchange, through Bluetooth, their polar coordinates and, according to 
the \textit{position negotiation protocol} described in Section
\ref{sec:pnp}, they adjust such coordinates
with the purpose to minimize the error of the mutual distance.
At the end of this protocol, $U_x$ obtains the correct pair $({\rho}_x, {\theta}_x)$.
Note that no pseudo-identifier of $U_x$ is exchanged in the negotiation protocol. Unlike classic BLE-based solutions, Bluetooth is used only to improve the accuracy of GPS. Moreover, another advantage of integrate Bluetooth in our solution is the following. In tradition GPS based solutions, when two users are close but separated by an obstacle, for example a wall, the server is not aware about this and registers a contact even if it does not happen. By using Bluetooth, the presence of the obstacle attenuates the signal and the contact is not captured by the smartphone. 

At this point, for the duration of the contact, $U_x$ sends, with frequency $\frac{1}{\tau}$, to a server $S$ (under the control of the health authority) the following information: $(h(C_i||R_P),R_x,{\rho}_x,{\theta}_x)$, where $h$ denotes a secure cryptographic hash function, $R_P$ is the current random sent by TSP to $U_x$, $R_x$ is the current random generated by $U_x$, and $({\rho}_x, {\theta}_x)$ are the adjusted polar coordinates of the position of $U_x$ respect to the centroid $C_i$.


Clearly, TSP or other entities must not be able to intercept the messages toward $S$. Therefore, the communication is encrypted by using the public key of $S$.

Note that $S$ is not able to recover the exact position of $U_x$ through the relative coordinates $({\rho}_x, {\theta}_x)$ because it cannot reverse the hash function in order to obtain the centroid $C_i$, thanks to the inclusion
of the \textit{salt} $R_P$. Indeed, without the random $R_P$, $S$ can perform a dictionary-based attack by testing all the possible centroids, whose number is always feasible for a brute-force attack.

Now, $S$ builds the tuple $(h(C_i||R_P),R_x,{\rho}_x,{\theta}_x, \tau_k)$
where $\tau_k=[t_k,t_{k+1}]$ denotes the $k$-th time-slot in which the information of $U$ arrives.
We use a time slot mechanism (instead of the absolute time) since two users simultaneously in the same microcell might not be perfectly synchronized to send their tuples. However, the time slots have not to be too large to avoid two users which enter in the microcell in different moments are be treated as they are in microcell simultaneously. 
We assume that $\tau=t_{k+1}-t_k$ for each $k$. In word, the size of the time slot is a constant value and coincides with the inverse of the frequency with which the users send their information to $S$.

In a later moment, $S$ searches all the tuples with values $(h(C_i||R_P),R_y,{\rho}_y,{\theta}_y, \tau_k)$ i.e., all the tuples sent by (possible) other users in the same microcell in the same time slot. Then, it computes, through the (adjusted) polar coordinates, the distance $d_{xy}$ between the users. 

For each of these tuples, $S$ searches, by using $(R_x,R_y)$ as key, an entry in the \textit{contact database} with values $(R_x, R_y, n_{xy}, D_{xy}, r_{xy})$ where $n_{xy}$ denotes the number of time-slot in which $U_x$ and $U_y$ came into contact with random $R_x, R_y$ respectively, $D_{xy}$ is the set of distances between them (for each the time-slot) and $r_{xy}$ is the \textit{partial risk level} computed as function of $D_{xy}$ and $n_{xy}$.
These entries are called \textit{contact bursts} since represent sequences, not necessarily consecutive, of contacts between two users.
If the contact burst between $R_x$ and $R_y$ exists:
\begin{enumerate}
    \item $n_{xy}$ is increased by one
    \item $d_{xy}$ is added to $D$
    \item $r_{xy}=f(D_{xy},n_{xy})$ is recomputed. 
\end{enumerate},  
Otherwise  (i.e., the contact burst does not exist), the entry $(R_x, R_y, n, D, r)$ is created with:
\begin{enumerate}
    \item $n_{xy}=1$
    \item $D_{xy}$ containing only $d_{xy}$
    \item $r_{xy}=f(D_{xy},n_{xy})$
\end{enumerate} 

We do not focus on the function for the computation of the partial risk level since it depends on several medical factors.
We can say that the function increases as the number of time-slot $n$ (i.e., the time interval) in which two users came into contact increases and it decreases as the distances between users increase.   
We just remark that all the information typically used to evaluate the risk in digital contact tracing solutions are available also in our model.


When a user $U_z$ tested positive for the infection in a health facility HF, she/he may choose to send her/his randoms to $S$. In order to avoid fake positive reports, we rely on a 1024 bits RSA blind signature scheme. As discussed in Section \ref{sec:security_analysis}, blind signature also avoids that, even though $S$ colludes with HF, it is not able to link all the randoms of $U_z$ to her/his real identity.
The procedure is the following. 
First, $U_z$ generates a random $A$ of 1024-256=
768 bits and obtains $M=A||h(A)$ where $h(A)$ is the application of a cryptographic hash function with digests of 256 bits (e.g., SHA-256). At this point, $U_z$ contacts HF to obtain the RSA blind signature on $M$. 
Let denote by $P$ the message with blind signature. $U_z$ unblinds $P$ and obtains the signature of HF $\sigma(M)$ of $M$. Finally, $U_z$ sends to $S$ $\sigma(M)$ and all the randoms $R_z$s she/he generated. 
$S$ verifies the signature and checks that $M=A||h(A)$.
In the positive case, $S$ searches all the contact burst including any of the randoms $R_z$s as first or second component. For each of these entries, $S$ sends in broadcast a pair containing the other random (i.e., the random generated by a user which is came into contact with $U_z$) and the partial risk level.

Each user $U_t$ receives a set of pairs $(R_i,r_i)$ and search the subset of pairs $P$ where $R_i$ coincides with any of his/her generated randoms $R_t$. If this subset is empty, $U_t$ has not encountered any infected users. Otherwise, she/he came into contact one or more times with one or more users. Finally, the partial risk levels occurring in the pairs of $P$ are combined together through another function which returns the \textit{total risk level} for the user $U_t$ (also the definition of this function is outside the scope of this paper).

\section{\uppercase{The ZE2-P3T position negotiation protocol}}\label{sec:pnp}

\begin{figure}[]
\centering
        {\includegraphics[width=6cm]{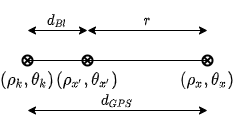}}
        \caption{The protocol PNP with $r>0$}
        \label{fig:PNP}
\end{figure}

In this section, we present a protocol performed by
the users to improve the accuracy of the coordinates captured through the GPS. This protocol involved pairs of users
and is named PNP, which stands for
\textit{position negotiation protocol}.
We say that a user is \textit{locked} if she/he has
executed PNP with another user, otherwise she/he is \textit{unlocked}.
Consider an unlocked user $U_x$ which enters in the action range of Bluetooth with a group $G$ of other users. For each locked user $U_y$ in $G$ with polar coordinates $({\rho}_y,{\theta}_y)$, $U_x$ retrieves such coordinates through Bluetooth. Since $U_y$ is locked, $({\rho}_y,{\theta}_y)$ are already \textit{adjusted}. On the other hand, the polar coordinates of $U_x$, $({\rho}_x,{\theta}_x)$, which are obtained through GPS, should be adjusted. 
To accomplish this, $U_x$ computes the distance $d_{GPS}=\sqrt{{{\rho}_x}^2+{{\rho}_y}^2-2{\rho}_x{\rho}_ycos({\theta}_x-{\theta}_y)}$ obtained by considering the non-adjusted coordinates of $U_x$.
Then, $U_x$, as typically done on the basis of the
signal power, computes again the distance with $U_y$. We denote by $d_{Bl}$ such a distance and assume it represents a more accurate estimate of $d_{GPS}$. Finally, among all (locked) users, $U_x$ chooses one of the users $U_k$ such that $|r|=|d_{GPS}-d{Bl}|$ is minimum, that is the user $U_k$ minimizing
the error of GPS w.r.t. Bluetooth
(which we can consider more accurate). 
If we denote by $({\rho}_k,{\theta}_k)$ the coordinates of $U_k$, the new coordinates of $U_x$, $({\rho}_x',{\theta}_x')$, are obtained by moving the old coordinates by $|r|$ along the straight line passing between $({\rho}_k,{\theta}_k)$ and $({\rho}_x,{\theta}_x)$ so that the new distance between $U_x$ and $U_k$ is equal to $d_{Bl}$, as depicted in Figure \ref{fig:PNP}.
 After this process, $U_x$ is locked.
 
 If no locked user exists in the action range of Bluetooth, for each (unlocked) user $U_y$, $U_x$ computes $d_{GPS},d_{Bl}$ and $r$, defined as above, and chooses a user $U_k$ with minimum value $|r|$. This time, both $U_x$ and $U_k$ update their coordinates, by moving them by $|r|/2$ along the straight line passing between $({\rho}_k,{\theta}_k)$ and $({\rho}_x,{\theta}_x)$ so that the new distance between $U_x$ and $U_k$ is equal to $d_{Bl}$.
 After this process, both $U_x$ and $U_z$ are locked.

Note that, as long as a user detect only another (locked or unlocked) user  through Bluetooth, our protocol works well. In fact, even if the adjusted coordinates are not necessarily correct, the distance between the two users is that measured through Bluetooth, which is widely considered acceptable for the purpose of proximity tracing. If more users participate in PNP, we use a greedy approach in order to minimize the adjusting of the coordinates and to obtain the Bluetooth distance at least with a user.

\begin{figure*}[t]
  \begin{center}
  \begin{tabular}{ ||c | c |  c | c|| }
        \hline
         Attack &  Low Cost DP-3T &  Unlinkable DP-3T &  ZE2-P3T  \\ \hline
         Paparazzi &  \xmark & \cmark & \cmark   \\ \hline
         Orwell &  \xmark & \xmark & \cmark   \\ \hline
         Brutus &  \xmark & \xmark & \cmark  \\ \hline
         Gossip &  \xmark & \xmark & \cmark  \\ \hline
         Matteotti &  \cmark & \xmark & \xmark  \\ \hline
         Missile &  \xmark & \xmark & \cmark  \\ \hline
         Fregoli & \xmark & \xmark & \cmark  \\ \hline
         Battleship &  \cmark & \cmark & \cmark  \\ \hline

     \end{tabular}
       \caption{Vulnerabilities of DP-3T and ZE2-P3T to the attacks. The symbol {\xmark } denotes that the solution is vulnerable while the symbol {\cmark } denotes that the solution resists to the attack.}
        \label{fig:table}
    \end{center}
     \end{figure*}

\section{\uppercase{Security Analysis}}
\label{sec:security_analysis}

The claimed robustness of the decentralized solutions like DP-3T mainly relies on the fact that identities are pseudo-random numbers that, as such, appear unlinkable to any observer. Unfortunately, this is true unless the seed from which these pseudo-randoms are generated is not known to the attacker.
What makes the linkability of identifiers a concrete privacy threat is that ephimerals identifiers are not kept only by the legitimate owner, but are exchanged among all the users.
As we will see in detail in the sequel of the section, the above possibility is realistic in both the designs of DP-3T (i.e., low-cost and unlinkable), under different attack models.
We will show that our solution is immune from this issue, just because no exchange of identifiers is enabled.

We analyse in detail the attacks on DP-3T known in the literature and show the above claim about our technique.

The involved actors of our security model are: 
\begin{itemize}
    \item The users $U$ which send, periodically, their randoms to the server $S$. If they find out to be infected, this information is reported to $S$.
    \item The server $S$ under the control of the health authority. It receives the randoms of the users and alerts them when a user communicates she/he is infected.
    \item The telephone service provider TSP which sends a random $R_P$ in a fixed area with several microcells, in order to prevent the server $S$ to identify the microcell where a user is located.
    \item The health facility HF which performs the tests on the users to diagnose the disease.
\end{itemize}

The attacker can be a generic entity (for example, a user or a company).
We assume that the health authority and TSP do not collude.
Consider that, in a real-life scenario, a collusion of the health authority with TSP (a private company) would easily come to light.

In the following, we show how our solution face the attacks discussed in \cite{avitabile2020towards} for which the DP-3T solution
is vulnerable plus some other relevant attacks. 

We highlight that many attacks are due to the exchange of the ephemeral identifiers among the users through Bluetooth. In our solution, no random is exchanged.

\noindent \textbf{Paparazzi Attack} \cite{avitabile2020towards}.
The attack aims to trace infected users by linking their ephemeral identifiers. We assume the server is trusted. This attack works only with the low-cost design of DP-3T. First, the attacker installs several passive BLE devices through the territory in order to collect the ephemeral identifiers of other users located in proximity of such devices. Moreover, it records the time and the location where such identifiers are received and, possibly, other information about the users. When a user $U_x$ results infected, she/he sends her/his secret key $SK$ to the server $S$ which, in turn, broadcasts it to all the users. Starting from $SK$, the attacker is able to generate all the ephemeral identifiers of $U_x$ and to track her/him through the information (time, location, etc.) stored when $U_x$ passed in proximity of the passive devices. Clearly, this attack does not work on the unlinkable design of DP-3T since the infected user $U_x$ sends the seeds to generate the ephemeral identifiers to the server $S$, but this latter does not broadcast such seeds to all users. Instead, $S$ generates all the ephemeral identifiers of $U_x$ and adds them to the Cuckoo filter, so that the attacker cannot link them.  
Similarly, also ZE2-P3T does not suffer of this kind of attack since the ephemeral identifiers, that are represented by the randoms generated by the users, are not exchanged, but are sent directly to the server.
Since the server is trusted, the attack cannot be performed.

\noindent \textbf{Orwell Attack} \cite{avitabile2020towards}.
The objective is the same as Paparazzi attack, but with the difference that the attacker colludes with the server $S$. Clearly, this time, also the unlinkable design of DP-3T is vulnerable to the attack since the server $S$ knows the seeds of the infected users and can easily link their ephemeral identifiers. We claim that, although, in principle, such an attack is possible in ZE2-P3T, it is definitely harder and less effective than in DP-3T. In fact, in order to know the randoms of users coming from a specific microcell, $S$ needs to know the random $R_P$ sent by TSP in that microcell. Since $S$ does not collude with TSP, the only way to obtain $R_P$ is to collaborate with a partner located in the area whenever $R_P$ is sent by TSP. To put on a mass tracking system, the attacker (colluding with the server) must have many partners spread throughout the territory and each one of them has to be registered with TSP to obtain $R_P$. Clearly, this is more onerous that to install passive BLE devices. Moreover, our solution includes in general a certain level of uncertainty, whenever more than one user belongs to a microcell simultaneously.

\noindent \textbf{Brutus attack}. 
In this attack, the health facility HF and the server $S$ collude to identify the mapping between pseudonymous and real identities of infected users. It is an exploit of the authorization mechanism with which infected users communicate their status to $S$.
DP-3T (both the designs) proposes three different authorization mechanisms but they are, essentially, based on an authorization code released by HF.
Clearly, HF knows the identity of the infected user and if it colludes with $S$, then it may provide to $S$ the mapping between the real identity of a user and its authorization code. $S$ can associate this identity with the ephemeral identifiers sent by the user. Both DP-3T designs are vulnerable to this attack. In ZE2-P3T, the authorization code is replaced by $M$ which cannot be linked by HF to the message submitted by the user to obtain the signature, thanks to the blind signature mechanism. Thus, both HF and $S$ cannot link $M$ to the real identity of the user. In conclusion, ZE2-P3T is not vulnerable to
Brutus attack.

\noindent \textbf{Gossip attack} \cite{avitabile2020towards}. 
The objective of this attack is to provide an evidence about an encounter with an infected user before the discovering of her/him positiveness to the infection. It can be view as a security flaw because it is a misuse of the system for an unintended scope, potentially threatening privacy and exploitable for disputes.
In both the designs (low-cost and unlinkable) of DP-3T, when the attacker captures the ephemeral identifiers of other users, she/he could, for example, store them on blockchain and, successively, demonstrate to have encountered such users. 
In ZE2-P3T, this attack is not possible since users do not exchange any random.

\noindent \textbf{Matteotti attack} \cite{avitabile2020towards}. 
In this attack, the objective is to deceive a user by providing her/him a fake contact with a positive user. The result aimed by the attacker is to damage the victim enforcing her/him quarantine (or other consequent actions). It requires the collusion of the attacker with the server. Suppose $U_v$ is the user victim of the attack. In the unlinkable design of DP-3T, the attacker places the BLE passive devices in proximity of $U_v$ and when this latter comes into contact with another user $U_s$, the passive devices capture the ephemeral identifiers of $U_s$ and send them to the server. The server inserts such identifiers in the Cuckoo filter so that, when $U_v$ checks the filter, she/he is wrongly alerted. 
Low Cost DP-3T is not vulnerable to this attack since the server is not able to generate the secret keys of the users starting from the collected ephemeral identifiers. 
Similarly to the unlinkable DP-3T, in ZE2-P3T, the server can notify false information about the contacts at risk.

Another attack with the same objective as the Matteotti attack is the following. It does not require the collusion with the server.

\noindent \textbf{Missile attack}. 
The objective of this attack is the same of the Matteotti attack. In this case, the attacker is a user who is positive to the disease. She/He can use a Bluetooth amplifier transmitter to send his/her ephemeral identifiers (like a \textit{missile}) to other users even very distant from her/him and so, not at risk. However, when the server communicates the infected identifiers of the attacker, such users are wrongly alerted. The attack is based on the exchange of the ephemeral identifiers through Bluetooth, so both the designs of DP-3T are vulnerable.
On the contrary, ZE2-P3T does not suffer from this attack since no identifier is exchanged through Bluetooth.

Another possible attack is the following.

\noindent \textbf{Fregoli attack}. 
This attack aims to simulate fake contacts between users. The attacker can collect the ephemeral identifiers of the users and use them in place of his/her own. This is then an impersonation attack, as its name evokes, being Fregoli one of the major \textit{quick-change} artists of the story.
The result of the impersonation is that a user $U_x$ keeps ephemeral identifiers of other users with which she/he never met. If any of them results infected, $U_x$ is wrongly alerted as in the Matteotti attack. This attack is more effective if the attacker uses a Bluetooth amplifier transmitter. Again, this attack is possible in both the designs of DP-3T but it is not possible in ZE2 -P3T since no random is exchanged through Bluetooth.

Finally, we conclude the analysis by presenting another attack which, potentially, affects GPS-based approaches.

\noindent \textbf{Battleship attack}. 
In this attack, the server tries to identify the position of the users to track them. In both the designs of DP-3T, such an attack is not possible since no information about the position is sent to $S$. On the contrary, any standard
GPS-based solution is affected by this problem. Therefore, it is important to check what happens for our protocol.
In ZE2-P3T, the user sends the polar coordinates relative to a given centroid $C_i$. Therefore, the attack would succeed if the server is able to identify such centroid. The user sends $h(C_i||R_P)$ and, even if the total number of centroids is not too large, the presence of the random $R_P$ makes dictionary-based attacks unfeasible. Since $S$ and TSP do not collude, the only way for $S$ would be to collaborate with a partner physically located in a microcell in order to obtain $R_P$. As explained in Orwell Attack, to put on a mass tracking system is infeasible.

We highlight that, although the attacks regard DP-3T, they shall also apply to others decentralized protocols \cite{tcn,mit-pact,chan2020pact} since the vulnerabilities are due to the exchange of the ephemeral identifiers.


Finally, we observe that, being our approach centralized, the intrinsic price we have to pay in terms of privacy is that, once an infected patient sends to the server her/him randoms used in the contagious window, the server links this randoms
(this is obviously necessary in the centralized model), learning some piece of pseudonym information about the user. We argue that as the match between real identities and pseudonyms is not possible even in case of collusion between HF and server (see Brutus attack above), this is not an actual threat to privacy, against the evident benefits given by our approach summarized in the table reported in Figure \ref{fig:table}.

\section{\uppercase{Conclusions}}
\label{sec:conclusion}
The fight against the pandemic of COVID-19 requires a number of coordinated actions that governments should take promptly. Among these, digital contact tracing has an important role, especially during the after-lock-down phase, in which potential infected people should be rapidly identified and isolated.
The main contribution of this paper is to show that a centralized approach,
exploiting GPS, can provide a solution definitely more effective, in terms of security and privacy protection, than decentralized solutions based on DP-3T or similar protocols. Unlike similar attempts occurring in the current literature, our solution does not rely on complex cryptographic mechanisms to avoid people position tracking, but only efficient cryptographic hashes and RSA blind signatures only for the reporting phase. 
As a future work we plan to implement the solution also by detailing the combination of existing technologies based on the Earth magnetic field to improve the indoor localization accuracy.
Another direction of further extension of this paper regards a more accurate (tested) definition of the function estimating the contagious risk, which is a task inherently interdisciplinary, outside the scope of this paper, aimed to rapidly share this new approach with the scientific community 
(also by publication in open-access pre-print archives), being the topic of high interest in the current days.

\bibliographystyle{IEEEtran}
\bibliography{bib}

\begin{thebibliography}{10}
\providecommand{\url}[1]{#1}
\csname url@samestyle\endcsname
\providecommand{\newblock}{\relax}
\providecommand{\bibinfo}[2]{#2}
\providecommand{\BIBentrySTDinterwordspacing}{\spaceskip=0pt\relax}
\providecommand{\BIBentryALTinterwordstretchfactor}{4}
\providecommand{\BIBentryALTinterwordspacing}{\spaceskip=\fontdimen2\font plus
\BIBentryALTinterwordstretchfactor\fontdimen3\font minus
  \fontdimen4\font\relax}
\providecommand{\BIBforeignlanguage}[2]{{%
\expandafter\ifx\csname l@#1\endcsname\relax
\typeout{** WARNING: IEEEtran.bst: No hyphenation pattern has been}%
\typeout{** loaded for the language `#1'. Using the pattern for}%
\typeout{** the default language instead.}%
\else
\language=\csname l@#1\endcsname
\fi
#2}}
\providecommand{\BIBdecl}{\relax}
\BIBdecl

\bibitem{cheng2020contact}
H.-Y. Cheng, S.-W. Jian, D.-P. Liu, T.-C. Ng, W.-T. Huang, and H.-H. Lin,
  ``Contact tracing assessment of covid-19 transmission dynamics in taiwan and
  risk at different exposure periods before and after symptom onset,''
  \emph{JAMA Internal Medicine}, 2020.

\bibitem{ferretti2020quantifying}
L.~Ferretti, C.~Wymant, M.~Kendall, L.~Zhao, A.~Nurtay, L.~Abeler-D{\"o}rner,
  M.~Parker, D.~Bonsall, and C.~Fraser, ``Quantifying sars-cov-2 transmission
  suggests epidemic control with digital contact tracing,'' \emph{Science},
  vol. 368, no. 6491, 2020.

\bibitem{gomez2012overview}
C.~Gomez, J.~Oller, and J.~Paradells, ``Overview and evaluation of bluetooth
  low energy: An emerging low-power wireless technology,'' \emph{Sensors},
  vol.~12, no.~9, pp. 11\,734--11\,753, 2012.

\bibitem{troncosodecentralized}
C.~Troncoso, M.~Payer, J.~Hubaux, M.~Salath{\'e}, J.~Larus, E.~Bugnion,
  W.~Lueks, T.~Stadler, A.~Pyrgelis, D.~Antonioli \emph{et~al.},
  ``Decentralized privacy-preserving proximity tracing. apr. 12, 2020. url:
  https://github. com/dp-3t/documents/raw/master,'' \emph{DP3T White
  Paper.pdf}.

\bibitem{apgo}
\BIBentryALTinterwordspacing
Apple and Google, ``Apple and google's exposure notification system,'' 2020.
  [Online]. Available: \url{https: // www.apple. com/ covid19/ contacttracing}
\BIBentrySTDinterwordspacing

\bibitem{avitabile2020towards}
G.~Avitabile, V.~Botta, V.~Iovino, and I.~Visconti, ``Towards defeating mass
  surveillance and sars-cov-2: The pronto-c2 fully decentralized automatic
  contact tracing system,'' Cryptology ePrint Archive, Report 2020/493, 2020.
  https://eprint. iacr.org, Tech. Rep., 2020.

\bibitem{dp3tblevulnerability}
\BIBentryALTinterwordspacing
``Privacy and security risk evaluation of digital proximity tracing systems.
  the dp-3t project,'' 2020. [Online]. Available:
  \url{https://github.com/DP-3T/Security analysis/Privacy and Security Attacks
  on Digital Proximity Tracing Systems.pdf}
\BIBentrySTDinterwordspacing

\bibitem{vaudenay2020centralized}
S.~Vaudenay, ``Centralized or decentralized?'' 2020.

\bibitem{EU2020}
E.~eHealth Network, ``Mobile applications to support contact tracing in the
  eu’s fight against covid-19 - common eu toolbox for member states,''
  \emph{versio 1.0}, 15.04.2020.

\bibitem{reichert2020privacy}
L.~Reichert, S.~Brack, and B.~Scheuermann, ``Privacy-preserving contact tracing
  of covid-19 patients,'' \emph{Sourced from}, 2020.

\bibitem{berke2020assessing}
A.~Berke, M.~Bakker, P.~Vepakomma, R.~Raskar, K.~Larson, and A.~Pentland,
  ``Assessing disease exposure risk with location histories and protecting
  privacy: A cryptographic approach in response to a global pandemic,''
  \emph{arXiv preprint arXiv:2003.14412}, 2020.

\bibitem{dar2020applicability}
A.~B. Dar, A.~H. Lone, S.~Zahoor, A.~A. Khan, and R.~Naaz, ``Applicability of
  mobile contact tracing in fighting pandemic (covid-19): Issues, challenges
  and solutions,'' Cryptology ePrint Archive, Report 2020/484, 2020,
  \url{https://eprint.iacr.org/2020/484}.

\bibitem{brack2020decentralized}
S.~Brack, L.~Reichert, and B.~Scheuermann, ``Decentralized contact tracing
  using a dht and blind signatures,'' \emph{Last accessed: 01st May}, 2020.

\bibitem{tcn}
\BIBentryALTinterwordspacing
``{TCN Protocol},'' 2020. [Online]. Available:
  \url{https://github.com/TCNCoalition/TCN}
\BIBentrySTDinterwordspacing

\bibitem{pepp}
\BIBentryALTinterwordspacing
P.-P. Team, ``Pan-european privacy-preserving proximity tracing,'' 2020.
  [Online]. Available: \url{https: // www. pepp-pt. org/}
\BIBentrySTDinterwordspacing

\bibitem{ntkpepp}
------, ``Pan-european privacy-preserving proximity tracing need-to-know
  system. overview of pepp-pt ntk,'' 2020.

\bibitem{aisec2020pandemic}
F.~AISEC, ``Pandemic contact tracing apps: Dp-3t, pepp-pt ntk, and robert from
  a privacy perspective,'' 2020.

\bibitem{altuwaiyan2018epic}
T.~Altuwaiyan, M.~Hadian, and X.~Liang, ``Epic: Efficient privacy-preserving
  contact tracing for infection detection,'' in \emph{2018 IEEE International
  Conference on Communications (ICC)}.\hskip 1em plus 0.5em minus 0.4em\relax
  IEEE, 2018, pp. 1--6.

\bibitem{singaporean}
\BIBentryALTinterwordspacing
D.~K. H.~Asghar, F.~Farokhi and B.~Rubinstein, ``On the privacy of
  tracetogether, the singaporean covid-19 contact tracing mobile app, and
  recommendations for australia,'' 2020. [Online]. Available:
  \url{http://tiny.cc/pb3lmz}
\BIBentrySTDinterwordspacing

\bibitem{blevul}
\BIBentryALTinterwordspacing
``Privacy and security risk evaluation of digital proximity tracing systems,''
  2020. [Online]. Available:
  \url{https://github.com/DP3T/documents/blob/master/Security analysis/Privacy
  and Security Attacks on Digital Proximity Tracing Systems.pdf}
\BIBentrySTDinterwordspacing

\bibitem{kampf2020persistence}
G.~Kampf, D.~Todt, S.~Pfaender, and E.~Steinmann, ``Persistence of
  coronaviruses on inanimate surfaces and its inactivation with biocidal
  agents,'' \emph{Journal of Hospital Infection}, 2020.

\bibitem{de2014indoor}
G.~De~Angelis, V.~Pasku, A.~De~Angelis, M.~Dionigi, M.~Mongiardo, A.~Moschitta,
  and P.~Carbone, ``An indoor ac magnetic positioning system,'' \emph{IEEE
  Transactions on Instrumentation and Measurement}, vol.~64, no.~5, pp.
  1267--1275, 2014.

\bibitem{fan2014cuckoo}
B.~Fan, D.~G. Andersen, M.~Kaminsky, and M.~D. Mitzenmacher, ``Cuckoo filter:
  Practically better than bloom,'' in \emph{Proceedings of the 10th ACM
  International on Conference on emerging Networking Experiments and
  Technologies}, 2014, pp. 75--88.

\bibitem{mit-pact}
\BIBentryALTinterwordspacing
P.~Team, ``Decentralized privacy-preserving proximity tracing,'' 2020.
  [Online]. Available:
  \url{https://pact.mit.edu/wp-content/uploads/2020/04/The-PACT-protocol-specification-ver-0.1.pdf}
\BIBentrySTDinterwordspacing

\bibitem{chan2020pact}
J.~Chan, D.~Foster, S.~Gollakota, E.~Horvitz, J.~Jaeger, S.~Kakade, T.~Kohno,
  J.~Langford, J.~Larson, P.~Sharma, S.~Singanamalla, J.~Sunshine, and
  S.~Tessaro, ``Pact: Privacy sensitive protocols and mechanisms for mobile
  contact tracing,'' 2020.

\end{thebibliography}

\end{document}